\documentstyle[preprint,aps,epsf]{revtex}

\newcommand{\be}{\begin{eqnarray}}
\newcommand{\ee}{\end{eqnarray}}
\begin{document} 

\begin{titlepage} 
\flushright{LBNL-42096}
\begin{center}
\Large {\bf Energy loss effects on heavy quark production
in heavy-ion collisions at $\sqrt{s} = 5.5$ $A$TeV\footnote{This work was 
supported in part by
the Director, Office of Energy Research, Division of Nuclear Physics
of the Office of High Energy and Nuclear Physics of the U. S.
Department of Energy under Contract Number DE-AC03-76SF00098.}
}
\end{center}
\begin{center}
{Ziwei Lin$^a$ and Ramona Vogt$^{a,b}$
}\\[2ex]
$^a$Nuclear Science Division, LBNL, Berkeley, CA 94720 \\
$^b$Physics Department, University of California at Davis, Davis, CA 95616
\\ [2em] 
\end{center}

\begin{abstract}
We study the effect of energy loss on charm and bottom quarks in high-energy
heavy-ion collisions including hadronization,
longitudinal expansion and partial 
thermalization. 
We consider in detail the detector geometry and single lepton energy cuts 
of the ALICE and CMS detectors at the Large Hadron Collider (LHC) to show
the large suppression of high $p_T$ heavy quarks and the consequences of 
their semileptonic decays.
\end{abstract}


\end{titlepage} 

\section{Introduction}

A dense parton system is expected to be formed in the early stage of
relativistic heavy-ion collisions due to
the onset of hard and semihard parton scatterings. Interactions among
the produced partons in this dense medium will most likely lead to 
partial thermalization and formation of a quark-gluon plasma. It is thus
important to study phenomenological signals of the early parton
dynamics, a crucial step towards establishing the existence
of a strongly interacting initial system and its approach to
thermal equilibrium. 

The energy loss of fast partons is a good probe of dense 
matter \cite{gw_jet} since a fast parton traversing the medium must 
experience multiple elastic collisions \cite{bt,tg} 
as well as suffer radiative energy loss\cite{gw_lpm,bdps,bdmps,mtdks}. 
In principle, the energy loss by a parton in medium, both by 
elastic\cite{bt,tg} and radiative\cite{gw_lpm,bdps,bdmps,mtdks} processes, is
independent of the quark mass in the infinite energy limit.  
At finite energies, studies show that the elastic 
energy loss has a weak mass dependence.  For example, in a medium with
$\alpha_s=0.3$, $n_f \approx 2.5$ and a temperature of 500 MeV, 
the elastic $dE/dx$ for
10 GeV charm and bottom quarks is $\approx$ $-1.5$ and $-0.5$ GeV/fm,
respectively \cite{mtdks}. The radiative loss is perhaps even more important.
Taking into account multiple scatterings 
and the Landau-Pomeranchuk-Midgal effect, the radiative energy loss of 
a fast massless quark, $dE/dx \simeq -3 \alpha_s <p_{T w}^2>/8$\cite{bdmps}, 
is controlled by the characteristic broadening 
of the transverse momentum squared of the parton, $<p_{T w}^2>$,
determined by the properties of the medium.  Recent estimates of the radiative
loss by charm and bottom quarks \cite{mtdks} suggest that the loss from this
source is much greater than the elastic loss, $dE/dx = -7.5$ and $-5$ GeV/fm
for 10 GeV charm and bottom quarks respectively.  The calculated loss depends 
on the initial energy of the parton and the density of the medium.  It is also
unclear precisely where these analytical results are applicable.  In this 
paper, we will assume a constant loss of $dE/dx = -1$
GeV/fm to study the phenomenology of energy loss on heavy quarks at the LHC.

Since heavy-flavored mesons carry most of the heavy quark energy after 
hadronization, the energy lost by heavy quarks traveling through the 
quark-gluon plasma is directly
reflected in the suppression of large $p_T$ heavy-flavored mesons.

Unfortunately, it is difficult to detect charm or bottom mesons directly
with current tracking technology because of the large number of produced 
particles in central $AA$ collisions. However, the invariant mass of the
lepton pairs from heavy quark decays is related to the relative momentum
of the $Q\overline{Q}$ pair, the dilepton yields in this region could 
become an indirect measurement of the heavy quark spectrum. Therefore, it
should be sensitive to the energy loss suffered by the heavy quarks
as they propagate through dense matter. 

In this paper, we examine the effects of heavy quark energy loss at LHC
energies, $\sqrt{s} = 5.5$ TeV for Pb+Pb collisions,
including hadronization of the heavy quarks,
longitudinal expansion and thermal fluctuations of the collision system, which 
are important for the dilepton spectrum from heavy quark decays.
At the LHC energy, the heavy quarks are produced at sufficiently
large $p_T$ for the hadronization mechanism to be important.
Because of the longitudinal expansion, the momentum loss in the longitudinal
direction is quite different from that in the transverse direction. 
Depending on the actual number of scatterings, the heavy quarks can escape the
system without energy loss or lose enough momentum to be
stopped entirely.  However, heavy quarks cannot be at rest 
in a thermal environment. In the most
extreme scenario when they are stopped, they must have a 
thermal momentum distribution in their local frame.
The resulting suppression of high invariant mass dileptons 
is then very sensitive to the phase space
restrictions imposed by the detector design. 

This paper is organized as follows. We explain our energy loss model
in section~\ref{sec-model}. In section~\ref{sec-hvq},
we discuss the effects of energy loss on the charm and bottom quark spectra 
and show the resulting dilepton spectra from
correlated heavy meson decays.   
To demonstrate the sensitivity to the phase space restriction, 
in section~\ref{sec-alice} we calculate the spectra of 
$e^+e^-,e^{\pm}\mu^{\mp}$ and $\mu^+\mu^-$ pairs from correlated heavy meson
pair decays
within the planned acceptances of the ALICE detector, taking into account the
detector geometry and single lepton energy cuts. The $\mu^+\mu^-$ spectra
within the CMS geometry is also calculated.
In section~\ref{sec-single} we calculate the single $e$ and $\mu$ 
spectra from charm and bottom decays within the ALICE and CMS acceptances.  
We summarize in section~\ref{sec-summary}.

\section{Modeling the Energy Loss}
\label{sec-model}

First, the phase space distribution of the heavy quarks and the space-time 
evolution of the dense matter must be specified.
In the Bjorken model \cite{bj}, the matter has a longitudinal 
fluid velocity $v^F_z=z/t$ in the
local frame, essentially the fluid velocity of
free-streaming particles produced at $z=0$ and $t=0$.
Transverse flow, which sets in later, is neglected and
both the medium and the heavy quarks are assumed to be produced at $z=0$, 
the same point at which expansion begins.
Then, for any space-time point, ($z,t$), a heavy quark
is in a fluid with the same longitudinal velocity.
In the fluid rest frame, the heavy quark thus has momentum $(0,\vec
p_T)$, reduced to $(0,\vec{p_T}^\prime)$ after energy loss.
Thus the momentum of the heavy quark changes
from $(m_T \sinh y, \vec p_T)$ to $(m^\prime_T \sinh y, 
\vec{p_T}^\prime)$ in the lab frame.
The heavy quark essentially loses its transverse momentum but retains its 
rapidity because it follows the longitudinal flow.

To simplify the calculations, spherical nuclei of radius 
$R_A = 1.2 A^{1/3}$ are assumed so that the transverse 
area of the system in central collisions
is the area of the nucleus, neglecting transverse expansion.
For a heavy quark with a transverse path, $l_T$, and mean-free path,
$\lambda$, in the medium, the average number of scatterings is 
$\mu=l_T/\lambda$.
The mean-free path is introduced to account for the finite probability of the
heavy quarks to escape the system without interaction or energy loss.
The actual number of scatterings, $n$, is generated from the Poisson
distribution, $P(n,\mu)=e^{-\mu} \mu^n/n!$.  
This corona effect is especially important for heavy quarks produced at the
edge of the transverse plane of the collision.  
In the rest frame of the medium, the heavy quark then experiences a total
momentum loss of $\Delta p = n \lambda \; dE/dx$.

When a heavy quark loses most or all
of its momentum in the fluid rest frame, it 
begins to thermalize with the dense medium.  
The heavy quark is considered to be 
thermalized if its final transverse momentum after energy loss, 
$p^\prime_T$, is smaller than the average transverse 
momentum of thermalized heavy quarks at temperature $T$.  
These thermalized heavy quarks have a random thermal momentum in the
rest frame of the fluid. The final momentum of the thermalized
heavy quark is obtained by transforming back from the local fluid frame 
to the center-of-mass
frame of the collision. The parameters used in the calculation 
are $dE/dx = -1$ GeV/fm, $\lambda=1$ fm and $T=150$ MeV.
Simulations at RHIC energies \cite{lvw}
suggest that once the heavy quarks are assumed to lose energy, significant 
suppression
of the heavy quark spectra appears as long as $|dE/dx| \geq \langle p_T \rangle
/R_A$ where $\langle p_T \rangle$ is the average transverse momentum of the
heavy quark which produces leptons inside the detector acceptance.  
{\it E.g.}\ at central rapidities with Pb beams and 
$\langle p_T \rangle = 3$
GeV,  the threshold energy loss is as low as
$\langle p_T \rangle/R_A \sim 0.4$ GeV/fm.

\section{Effects of Energy Loss on Heavy Quark Production and Decay}
\label{sec-hvq}

The momentum distribution of the $Q \overline Q$ pairs is generated
from PYTHIA 6.115 \cite{jetset}.
Initial and final state radiation effectively simulates
higher-order contributions to heavy quark production so that the pair is no
longer azimuthally back-to-back as at leading order.  The
MRS D$-^\prime$ \cite{mrsd-p} parton distribution functions are used to
normalize the charm pair production cross section to
17.7 mb in $pp$ collisions at $\sqrt{s} = 5.5$ GeV \cite{vogt2}.  
The number of $Q \overline Q$ pairs in a Pb+Pb collision at impact parameter $b
= 0$ is obtained by multiplying the $pp$ production cross section by the
nuclear overlap function,
\be N_{Q \overline Q} = \sigma_{Q \overline Q}^{pp} T_{\rm PbPb}({\bf 0}) \ee
where $T_{\rm PbPb}({\bf 0}) = 30.4$/mb.  This scaling results in
540 charm pairs in a central Pb+Pb event. 
The $b \overline b$ production cross section is 224 $\mu$b
in $\sqrt{s} = 5.5$ TeV $pp$ collisions, leading to 
6.8 $b\overline b$ pairs in central Pb+Pb events. 
Although, as pointed out in the introduction, the bottom quark energy loss 
may be different from that of charm
quarks, the same parameters are used.  

Only dileptons from correlated 
$Q \overline Q$ pair decays, $N_{ll}^{\rm corr} = N_{Q \overline
Q}B^2(Q/\overline Q \rightarrow l^\pm X)$ are considered, 
{\it i.e.}, a single $Q \overline Q$ pair produces the dilepton.  
Dileptons from uncorrelated $Q \overline Q$ 
decays, which appear at higher invariant mass than those from correlated 
decays due to their larger rapidity gap, will
be particularly abundant for charm decays since $N_{ll}^{\rm uncorr} = N_{Q 
\overline Q}(N_{Q \overline Q}-1)B^2(Q/\overline Q \rightarrow l^\pm X)$. The
finite acceptance of a real detector will significantly reduce the uncorrelated
rate and like-sign 
subtraction should remove most of the remainder.  In practice
however, full subtraction will be difficult.  Another problem arises from
uncorrelated lepton pairs from a heavy quark and a background $\pi$ or $K$
decay.  Treatment of these uncorrelated backgrounds is not considered in  
this work.

Since the dilepton spectra in the LHC detectors 
are sensitive to decays of charm quarks with $p_T > 20$
GeV, the charm spectrum was generated in two steps to obtain a sufficient
number of high $p_T$ charm quarks.  First $10^5$ normal $c \overline c$ pairs
were generated followed by an equal number of $c \overline c$ pairs with a high
$p_T$ trigger such that the $c \overline c$ pair spectrum contains pairs with
$p_{T,c} > 5$ GeV and $p_{T, \overline c} > 5$ GeV only.  These high $p_T$ $ c
\overline c$ pairs were then removed from the normal spectrum
so that the resulting soft $c \overline c$ spectrum contains those pairs with
$p_{T,c} < 5$ GeV or $p_{T, \overline c} < 5$ GeV.  The relative weight of the
high $p_T$ spectrum is obtained from the ratio of the high $p_T$ events to the
total spectrum.  Because the bottom quarks have a harder $p_T$ spectrum
than the charm
quarks, such a proceedure was unnecessary for $b \overline b$ pairs.

\begin{figure}[h]
\setlength{\epsfxsize=\textwidth}
\setlength{\epsfysize=0.6\textheight}
\centerline{\epsffile{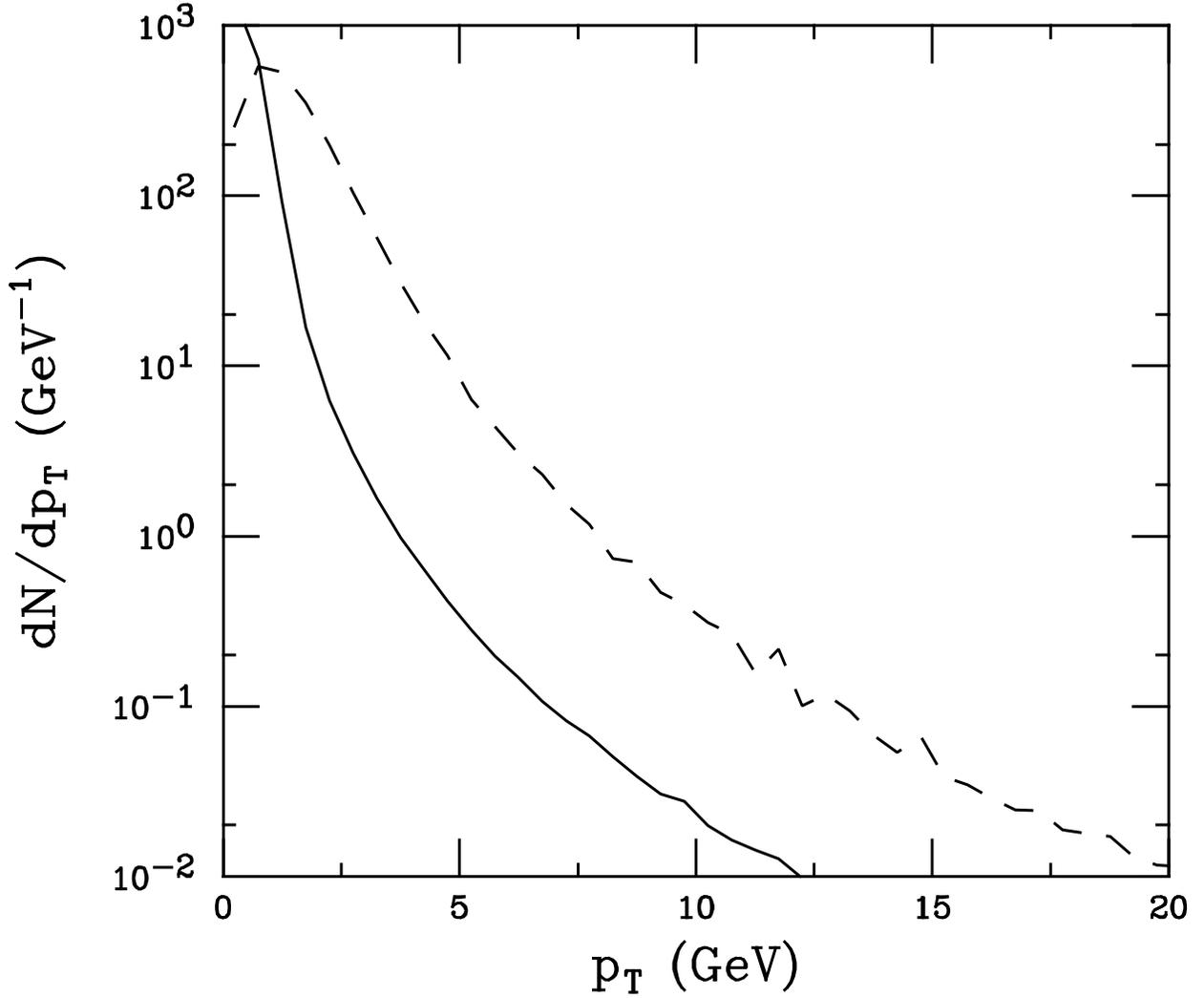}}
\vspace{1cm}
\caption{ The $p_T$ distribution of single $D$ mesons, integrated over all
phase space. The dashed curve is
without energy loss, the solid curve includes energy loss with $dE/dx =
-1$ GeV/fm.
}
\label{pt_c}
\end{figure}

In Fig.~\ref{pt_c} the single $D$ meson $p_T$ distribution is
shown without any phase space cuts.  The spectra in Fig.~\ref{pt_c} are
normalized, as are all the figures, to a single Pb+Pb event.
The dashed curve shows the generated
spectra without energy loss while the solid curve is the distribution after
energy loss.  Thermalization of charm quarks that have lost most of their
momentum causes the build-up at low $p_T$.  At
higher values of $p_T$, some quarks are sufficiently
energetic to escape the dense medium without being thermalized. For $p_T \geq
5$ GeV, the energy loss causes the $p_T$ distribution to drop nearly an order
of magnitude.

\begin{figure}[h]
\setlength{\epsfxsize=\textwidth}
\setlength{\epsfysize=0.6\textheight}
\centerline{\epsffile{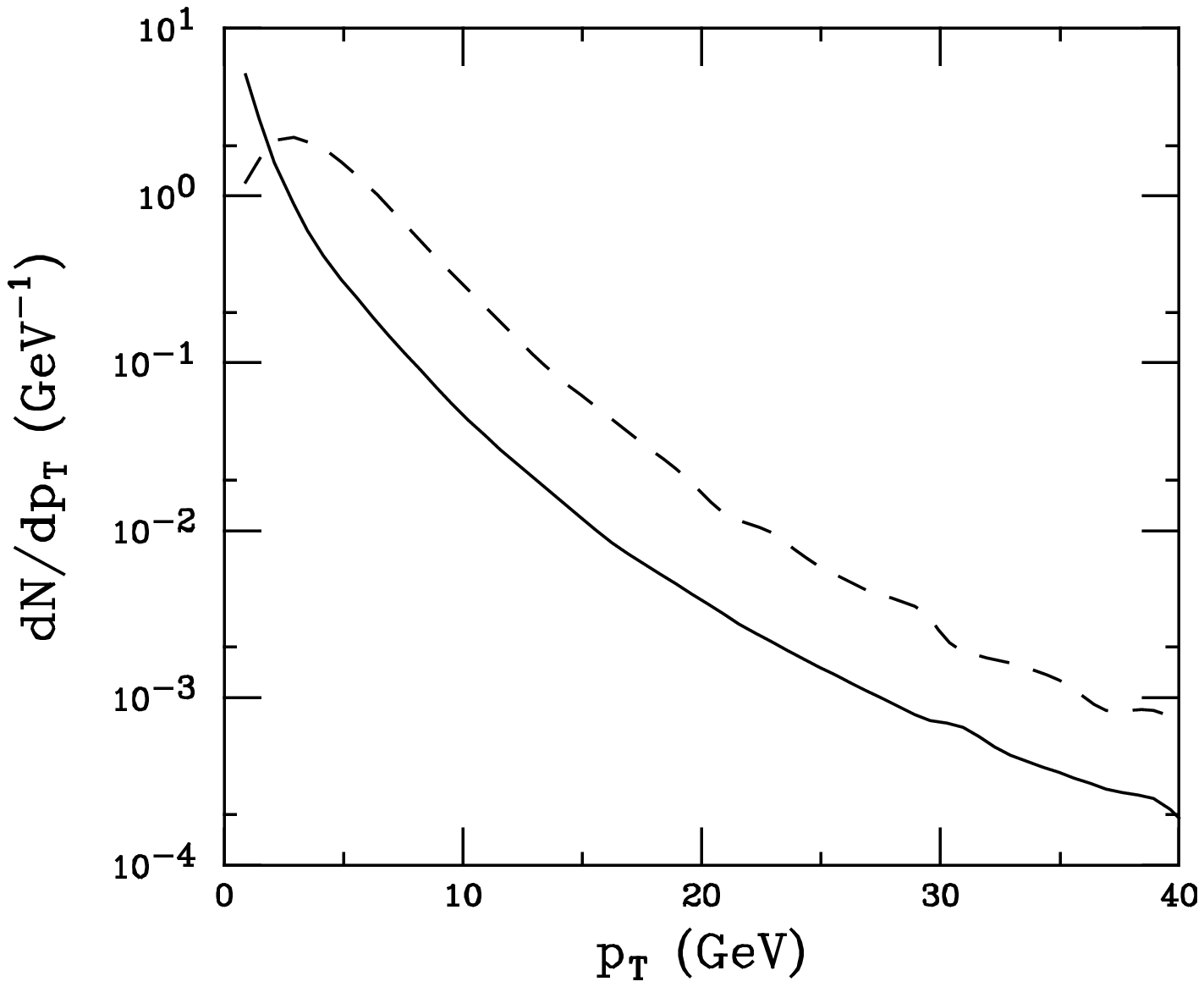}}
\vspace{1cm}
\caption{ The $p_T$ distribution of single $B$ mesons, integrated over all
phase space. The dashed curve is
without energy loss, the solid curve includes energy loss with $dE/dx =
-1$ GeV/fm.
}
\label{pt_b}
\end{figure}

Figure \ref{pt_b} shows the corresponding single bottom $p_T$ distribution.
The same trends are seen for bottom as in Fig.~\ref{pt_b} 
except that the energy
loss results in only a factor of five reduction in the high $p_T$ bottom yield.

In order to obtain the final meson distributions, 
the heavy quark distributions are convoluted with a fragmentation function.
While a delta-function type of fragmentation is suficient for low $p_T$
hadroproduction \cite{delfrag,e769}, high $p_T$ heavy quarks should fragment
according to a Peterson-type function \cite{Pete}, $D(z) \propto [z(1 - 1/z -
\epsilon/(1-z))^2 ]^{-1}$ where $z = p_H/p_Q$ and
$\epsilon = 0.06$ for charm and 0.006 for
bottom, determined from $e^+e^-$ interactions \cite{chrin}.  
Note that the heavy quark quantities are denoted by $Q$
while the heavy hadron formed from the fragmentation of the quark is denoted
with $H$. A corresponding intrinsic $k_T$ kick of 1 GeV for the partons in the
proton is also included.  Because fragmentation reduces the momentum of the
heavy-flavored hadron relative to the heavy quark, especially for charm, the
rapidity distribution of the final-state hadron is modified relative to the
fluid so that the hadron does not precisely follow the longitudinal flow.  In a
high-energy collision, $\sqrt{s}/m \gg 1$, the heavy quark rapidity
distribution is essentially flat.  However, the hadronization of the heavy 
quark enhances the rapidity distribution at central rapidities.  If the
delta-function type of fragmentation is assumed, the momentum does not change,
$p_Q = p_H$, but $E_Q^2 = E_H^2
- m_H^2 + m_Q^2$, resulting in a rapidity shift
\be dn \propto
dy_Q = \frac{dp_{zQ}}{E_Q} =  \frac{dp_{zH}}{E_Q} = \frac{m_{T,H} \cosh y_H
dy_H}{\sqrt{m_{T,H}^2 \cosh^2 y_H - m^2_H + m_Q^2}} \approx \frac{\cosh y_H
dy_H}{\sqrt{\cosh^2 y_H - \alpha^2}} \ee
where \be \alpha^2 = \frac{m_H^2 - m_Q^2}{m_{T,H}^2} \, \, . 
\label{alffrag} \ee
For $m_c = 1.3$ GeV, $m_D = 1.87$ GeV and $m_{T,D} \approx \sqrt{2} m_D$,
$\alpha^2 = 0.25$, enhancing the $D$ distribution at $y_H = 0$ by
$\approx 15$\%.  When $m_b = 4.75$ GeV, $m_B = 5.27$ GeV and $m_{T,B} =
\sqrt{2}m_B$, $\alpha^2 = 0.09$, enhancing the $B$ distribution by 
$\approx 5$\%.  The range of the
enhancement is $|y_H| < 2.5$.  If the Peterson function is used instead,
$\alpha^2$ increases,
\be \alpha^2 = \frac{m_H^2 - z^2 m_Q^2}{m_{T,H}^2} \,\, , \label{alfpete} \ee
increasing the $D$ enhancement at $y_H=0$ to $\approx 30$\% for $\langle z
\rangle \approx 0.7$ and the $B$ enhancement to $\approx 15$\% for $\langle z
\rangle \approx 0.85$.  These $\langle z \rangle$ values are typical for the
Peterson function with $\epsilon$ values given above.
The fragmentation thus tends to pile-up heavy hadrons at central rapidities.

The
dilepton spectrum from semileptonic charm and bottom decays may be used
to indirectly measure heavy quark 
production when a direct measurement via tracking is difficult.
Measurements of high-mass dileptons are themselves important.  
Copious thermal dilepton production\cite{thermal} was proposed as 
a signal of the formation of a thermally and chemically equilibrated
quark-gluon plasma. In order to obtain the thermal 
dilepton yields, the background from heavy quark decays must be subtracted.  
When energy loss was not included, dileptons from open charm decays 
at RHIC were shown to be about an order of magnitude higher than 
the contributions from the Drell-Yan process and bottom decays\cite{vogt2,shv},
making them the dominant background to the proposed thermal dileptons.
This background was determined to be even higher at the LHC \cite{vogt2}.
Energy loss changes the heavy quark momentum distribution as well as  
the resulting dilepton spectra from heavy quark decays.
Therefore, understanding the effect of energy loss on dileptons 
from heavy quark decays could also be an important step towards the 
observation of thermal dilepton signals.

The average branching ratios of
$\overline D \rightarrow l X$ are $\approx 12$\%. 
The lepton energy spectrum from $D$ meson semileptonic decays in
PYTHIA 6.115 is consistent with the measurement of the MARK-III
collaboration \cite{mark3}.  The $b$ quarks are assumed to  
fragment into $B^-, \overline B^0, \overline B^0_s$ 
and $\Lambda_b^0$ with production percentages 38\%, 38\%, 11\% 
and 13\%, respectively. Single leptons from bottom decays can be 
categorized as primary and secondary leptons.  
Leptons directly produced in the decay $B \rightarrow l X$ are
primary leptons while those indirectly produced, $B \rightarrow D X \rightarrow
l Y$, are secondary.  
Primary leptons have a harder energy spectrum than secondary leptons.  
A decaying $b$ hadron mainly produces primary
$l^-$ and secondary $l^+$ although 
it can also produce a smaller number of
primary $l^+$ due to $B^0-\overline B^0$ mixing.  The branching ratios of
the necessary bottom hadron decays are 9.30\% to primary $l-$, 2.07\% to 
secondary $l-$, 1.25\% to primary $l^+$, and 7.36\% to secondary $l^+$.  
The total number of dileptons from a 
$b\overline b$ decay can be readily estimated to be 0.020.  
Another important source of dileptons from bottom decays is
the decay of a single bottom, $B \rightarrow D l_1 X \rightarrow l_1 l_2 Y$.
The branching ratio for a single $B$ meson to a dilepton is 0.906\%,
therefore this source gives 0.018 dileptons,
comparable to the yield from a $b \overline b$ pair decay.
These branching ratios \cite{PDG}
and energy spectra from PYTHIA 6.115, consistent with
measurements \cite{bdecay}, are almost identical for muons and electrons.

\begin{figure}[h]
\setlength{\epsfxsize=\textwidth}
\setlength{\epsfysize=0.6\textheight}
\centerline{\epsffile{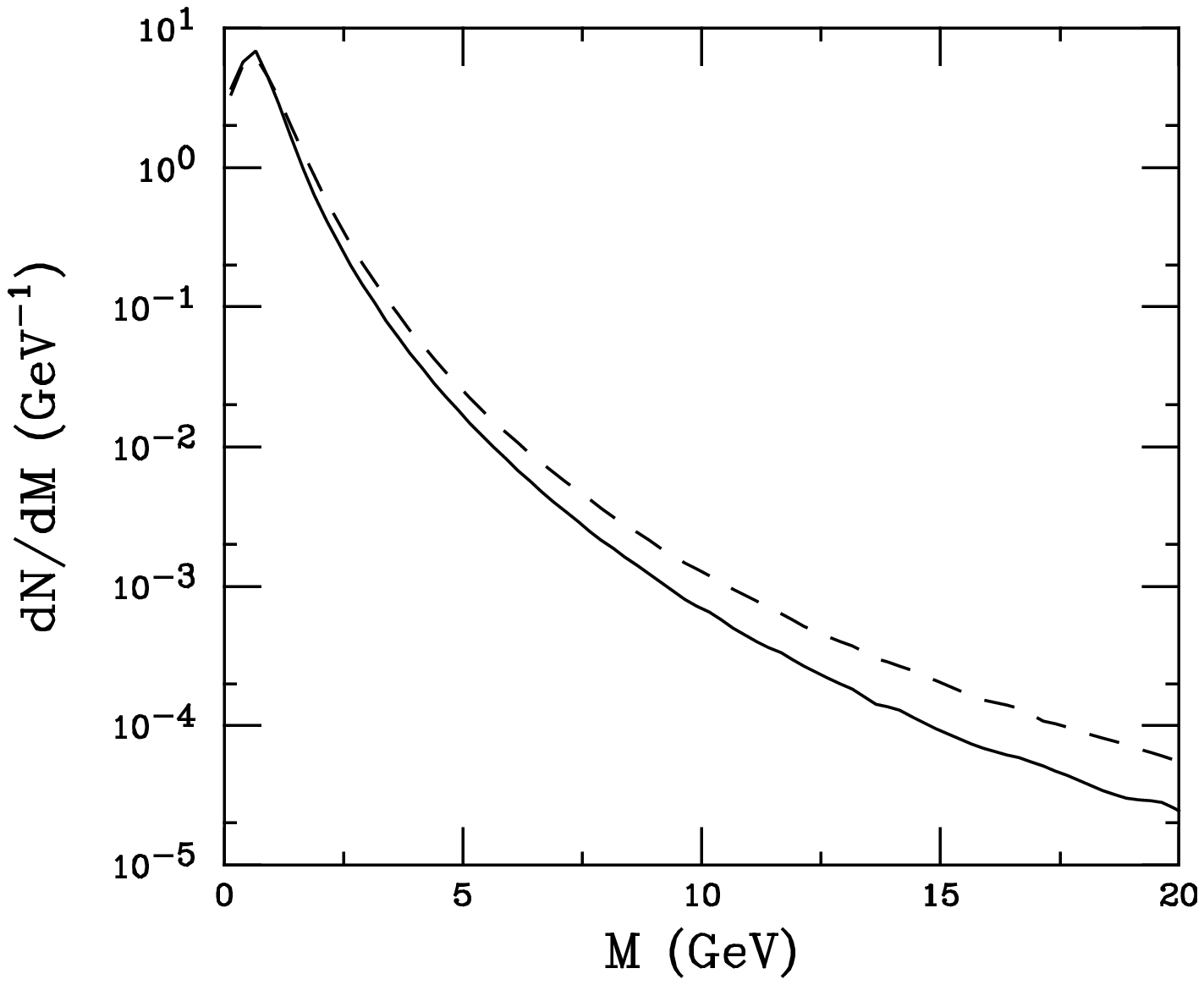}}
\vspace{1cm}
\caption{ The invariant
mass distribution of lepton pairs from correlated $D \overline D$ decays,
integrated over all phase space.  The dashed curve is
without energy loss, the solid curve includes energy loss with $dE/dx =
-1$ GeV/fm.
}
\label{m_cll}
\end{figure}

The dilepton invariant mass and rapidity are defined as:
\be
M &=&\sqrt {(p^\nu_{l^+}+p^\nu_{l^-})^2} \nonumber \\
y &=&\tanh ^{-1} \frac {p_{l^+}^z +p_{l^-}^z} {E_{l^+} +E_{l^-}}
\, \, . \ee
In Fig.~\ref{m_cll} the
dilepton invariant mass spectrum from correlated $D \overline D$ decays are
shown without any phase space cuts.
The dashed curves are the generated
spectra without energy loss while the solid curves are the distributions after
energy loss.  
Except for the small difference between the electron and 
muon masses, this spectrum represents both dielectrons and dimuons while the 
spectrum of opposite-sign $e\mu$ is a factor of two larger $(e^+\mu^- +
e^-\mu^+)$.  

\begin{figure}[h]
\setlength{\epsfxsize=\textwidth}
\setlength{\epsfysize=0.6\textheight}
\centerline{\epsffile{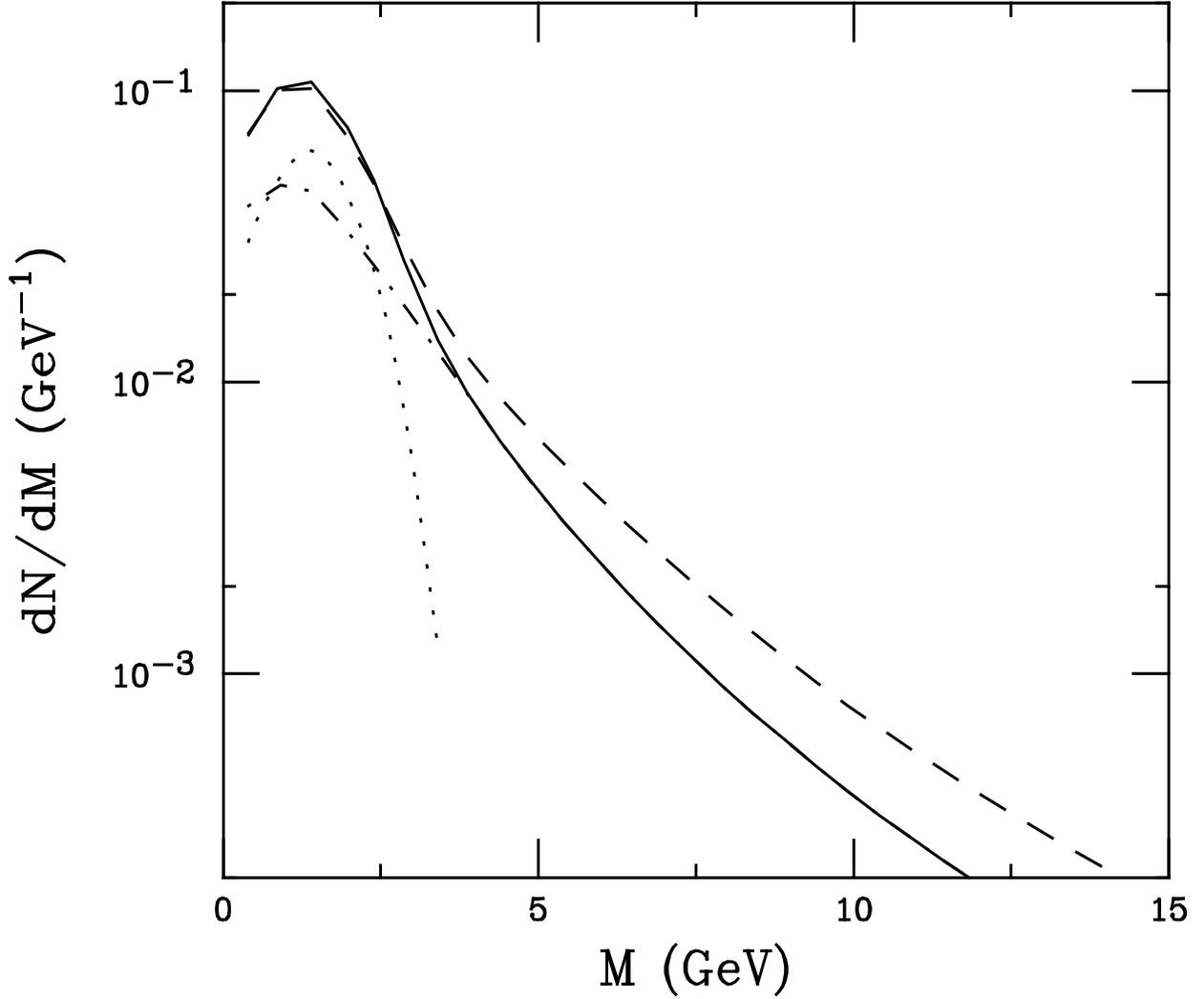}}
\vspace{1cm}
\caption{ The invariant
mass distribution of lepton pairs from correlated $B \overline B$ decays and
single $B$ decays, integrated over all phase space.  
The dotted curve is the contribution from semileptonic decay chains of single
$B$ mesons while the dot-dashed curve is from correlated $B \overline B$
decays.  Both include energy loss. The dashed curve is 
without energy loss and the solid curve includes energy loss with $dE/dx =
-1$ GeV/fm.  Note that the dashed and solid curves include all single
$B$ and $B \overline B$ pair decays.
}
\label{m_bll}
\end{figure}

Figure \ref{m_bll} shows
the integrated invariant mass spectra from correlated $B \overline B$ and 
single $B$ decays.  The dotted curve is the result of the decays of a single
$B$ to lepton pairs.  When $M < 3$ GeV, this contribution is larger than the
dilepton yield from $B \overline B$ decays, shown in the dot-dashed curve.
Both include energy loss.  The solid curve is the sum of the two contributions
while the dashed curve is the sum of single and pair decays to dileptons
without energy loss.  The same trends are seen for bottom as well as charm
except that the suppression of the spectrum due to energy loss begins at larger
invariant mass.  The mass distribution in
Fig.~\ref{m_bll} is truncated to more clearly show the contribution from
single $B$ decay.

\begin{figure}[h]
\setlength{\epsfxsize=\textwidth}
\setlength{\epsfysize=0.6\textheight}
\centerline{\epsffile{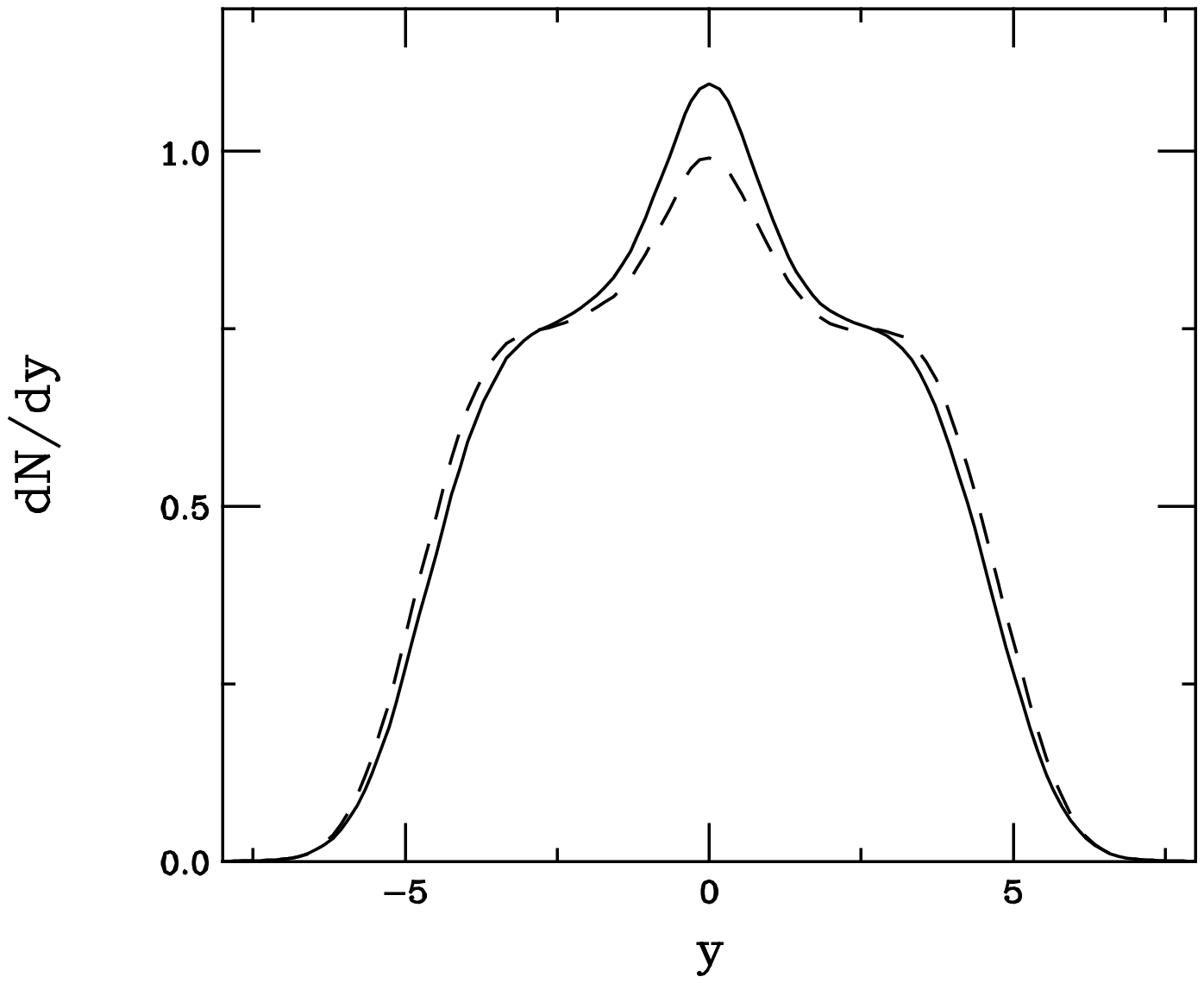}}
\vspace{1cm}
\caption{ The rapidity
distribution of lepton pairs from correlated $D \overline D$ decays,
integrated over all phase space.  The dashed curve is
without energy loss, the solid curve includes energy loss with $dE/dx =
-1$ GeV/fm.
}
\label{y_cll}
\end{figure}

In Fig.~\ref{y_cll}, the lepton pair rapidity distribution from correlated $D
\overline D$ decays with and without energy loss is shown.  The spectrum
reflects the effect of the hadronization shown in
eqs.~(2)-(4).  

\begin{figure}[h]
\setlength{\epsfxsize=\textwidth}
\setlength{\epsfysize=0.6\textheight}
\centerline{\epsffile{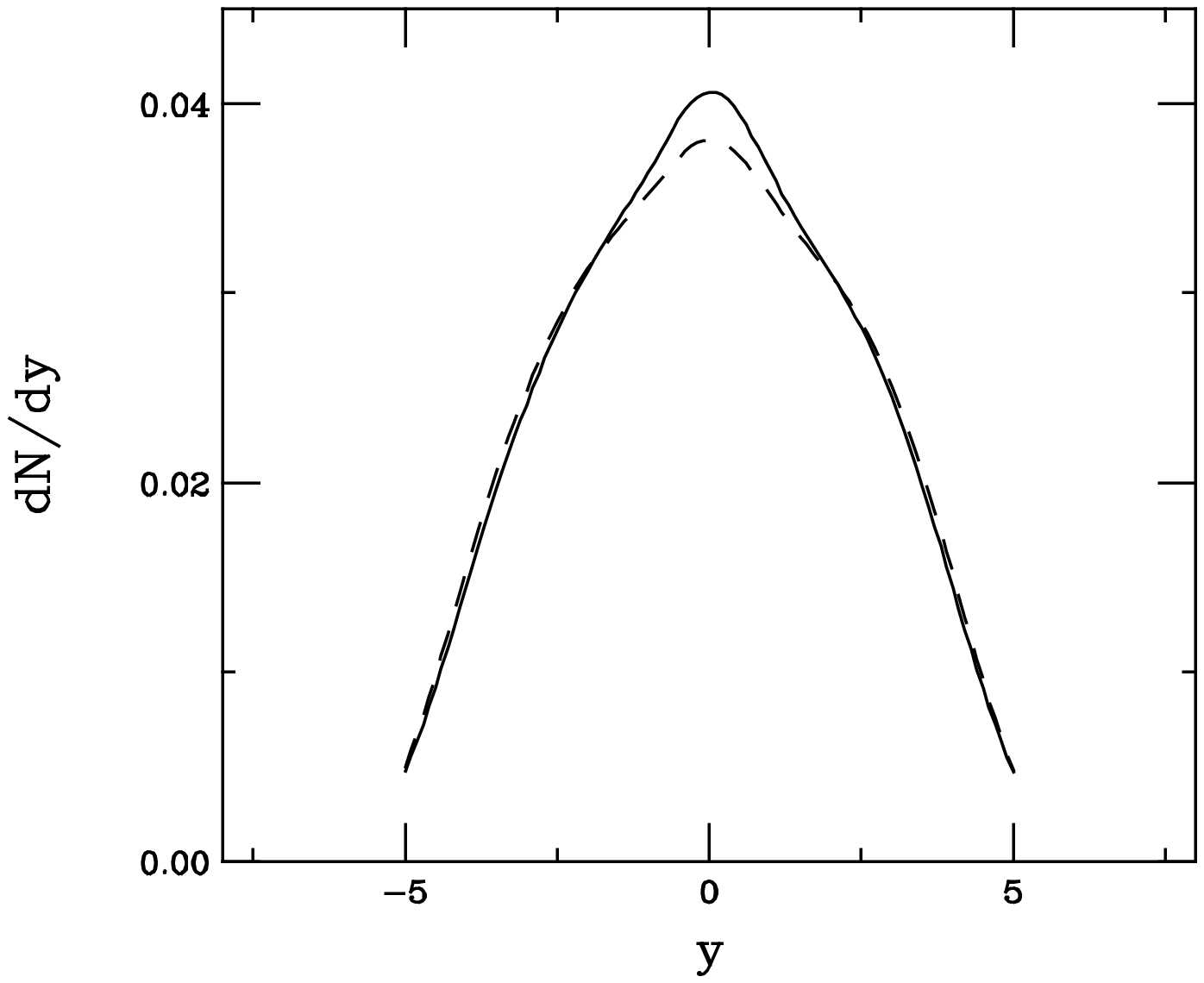}}
\vspace{1cm}
\caption{ The rapidity
distribution of lepton pairs from correlated $B \overline B$ and single $B$
decays, integrated over all phase space.  The dashed curve is
without energy loss, the solid curve includes energy loss with $dE/dx =
-1$ GeV/fm.
}
\label{y_bll}
\end{figure}

The lepton pair rapidity distribution from $B \overline B$ and single $B$
decays shows a similar shift in Fig.~\ref{y_bll}.  Since the $B \overline B$
decay distribution is not as broad as the $D \overline D$ distribution due to
the higher mass $B \overline B$ pairs, the narrowing of the central peak seen
in Fig.~\ref{y_bll} is not as dramatic as the charm hadron decays in
Fig.~\ref{y_cll}. 

A comparison of the dilepton spectra before and after energy loss
would naively suggest that the overall effect is small.  However, this
impression is misleading because the spectrum is integrated over the entire
phase space.  Heavy quarks and antiquarks in a pair tend to be separated by a
significant rapidity gap.  This gap can cause the invariant mass of the
subsequent lepton pair to be large.  However, once the finite detector
geometries are included, the effect of energy loss becomes more dramatic
as we show in the following section.

\section{Dileptons from Heavy Quark Decays in Real Detectors}
\label{sec-alice}

\subsection{ALICE}

The ALICE detector \cite{alice} consists of a central barrel with electron
detection capability and a forward muon
arm.  Thus it is well suited to carry 
out a comparative study of single lepton ($e,\mu$) and dilepton 
($ee, e\mu, \mu\mu$) yields comparable to the electromagnetic capabilities of
the PHENIX detector \cite{cdr} at RHIC.
(Note that PHENIX has two muon arms.) 
In this section, we calculate the dilepton yields 
within the designed ALICE acceptance.  

The ALICE central barrel covers $\pm 45^\circ$, corresponding to $| \eta | <
0.9$, with full azimuthal acceptance.  The forward muon arm covers the
polar angle $2^ \circ \leq \theta_\mu \leq
10^\circ$, corresponding to the pseudo-rapidity interval $2.5 \leq
\eta_\mu \leq 4$, again with full azimuthal coverage.  
We take $p_{T,e} > 1$ GeV and $p_{T, \mu} > 1$ GeV to
reduce the lepton backgrounds from random hadron decays.

\begin{figure}[h]
\setlength{\epsfxsize=\textwidth}
\setlength{\epsfysize=0.6\textheight}
\centerline{\epsffile{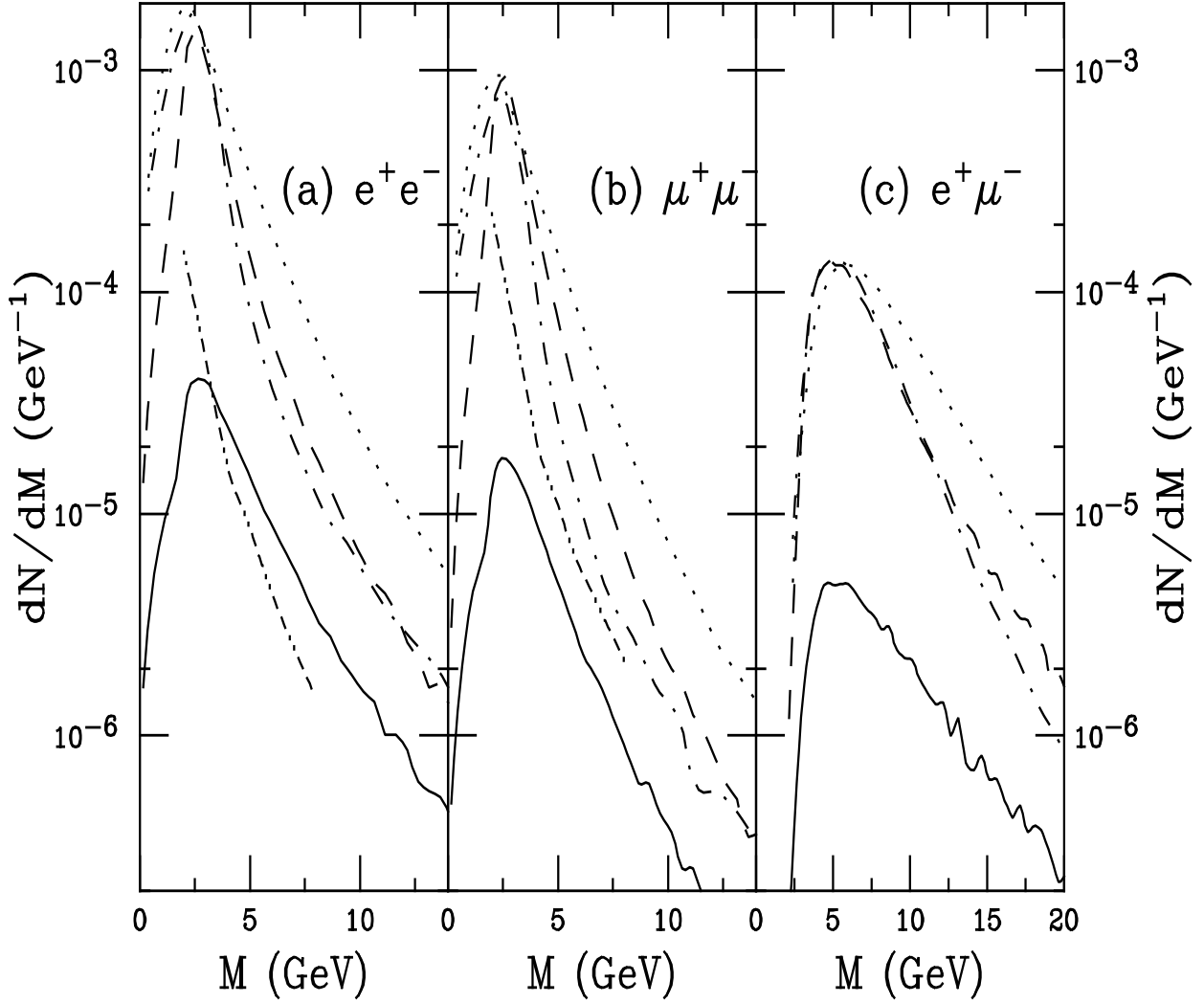}}
\vspace{1cm}
\caption{The dilepton invariant mass distributions in the ALICE acceptance.
The $e^+ e^-$ (a), $\mu^+ \mu^-$ (b) and $e \mu$ (c) channels are shown.
The dashed and dotted curves are the $D \overline D$ and summed single $B$ and
$B \overline B$ decays respectively without energy loss.  The solid and 
dot-dashed curves are the corresponding results with $dE/dx = -1$ GeV/fm.  
The Drell-Yan rate is given by the dot-dot-dashed curve in (a) and (b).
}
\label{malice}
\end{figure}

Fig.~\ref{malice} shows the invariant mass distribution of three types of
dileptons from open charm and bottom hadron
decays within the ALICE acceptance.  
The $e\mu$ spectrum includes both $e^+\mu^-$ and $e^-\mu^+$.  
From the comparison of our energy loss results with the initial distributions, 
we note that the three dilepton yields from charm decays and also for bottom
decays have rather similar suppression
factors.  This is different from the effect expected at RHIC because we use the
same $p_T$ cut for electrons and muons in ALICE while the electron and muon
energy cuts are the same in PHENIX \cite{cdr}.
Note the similarity of the charm and bottom hadron decay rates without energy 
loss for $M< 5$ GeV despite the much larger $c \overline c$
production rate.
Although there is significant suppression due to energy loss at high
invariant mass, the peaks of the bottom decay
spectra are not strongly suppressed.  

\begin{figure}[h]
\setlength{\epsfxsize=\textwidth}
\setlength{\epsfysize=0.6\textheight}
\centerline{\epsffile{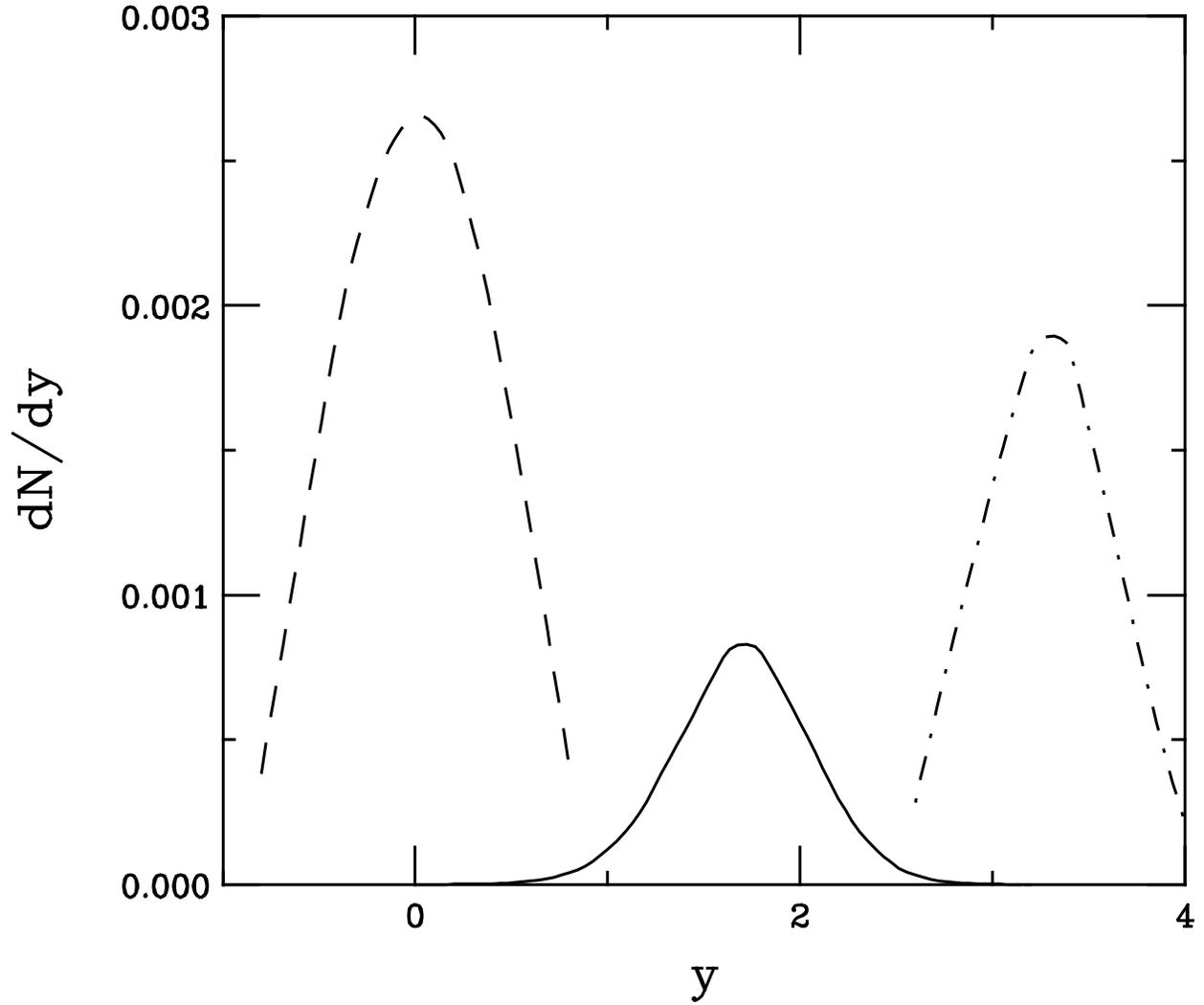}}
\vspace{1cm}
\caption{ The rapidity
distribution of lepton pairs from correlated $D \overline D$ decays
in the ALICE acceptance without energy loss.  The $e^+ e^-$ (dashed), $e \mu$
(solid) and $\mu^+ \mu^-$ (dot-dashed) acceptances are shown.
}
\label{yalice}
\end{figure}

To demonstrate the acceptance of the ALICE detector, in 
Fig.~\ref{yalice} we show the rapidity distribution of the three types of
dileptons from charm meson decays.  The $ee$ pairs are
centered  around $y \sim 0$, while the $e\mu$ acceptance covers 
pair rapidity around 1 to 2.5 
and the $\mu\mu$ pairs are found with $y \sim$ 2.5 to 4. 

We also plot the
Drell-Yan yields of dielectrons and dimuons in Fig.~\ref{malice}. 
After energy loss, the Drell-Yan dileptons
are greater than the open charm meson decays above 5 GeV in the $ee$ channel 
and
everywhere in the $\mu \mu$ channel.  However, the effect of energy loss in the
model is sufficiently weak for the $B \overline B$ and single $B$ decays to
remain above the Drell-Yan rate.  While thermal dileptons remain essentially
unobservable at the LHC, the suppression of dileptons from bottom hadron decays
is similar enough to present a good opportunity to measure the energy loss. 
Also note that high-mass $e\mu$ pairs cannot be used for charm measurements but
for bottom observation.

\subsection{CMS}
\label{sec-cms}

\begin{figure}[h]
\setlength{\epsfxsize=\textwidth}
\setlength{\epsfysize=0.6\textheight}
\centerline{\epsffile{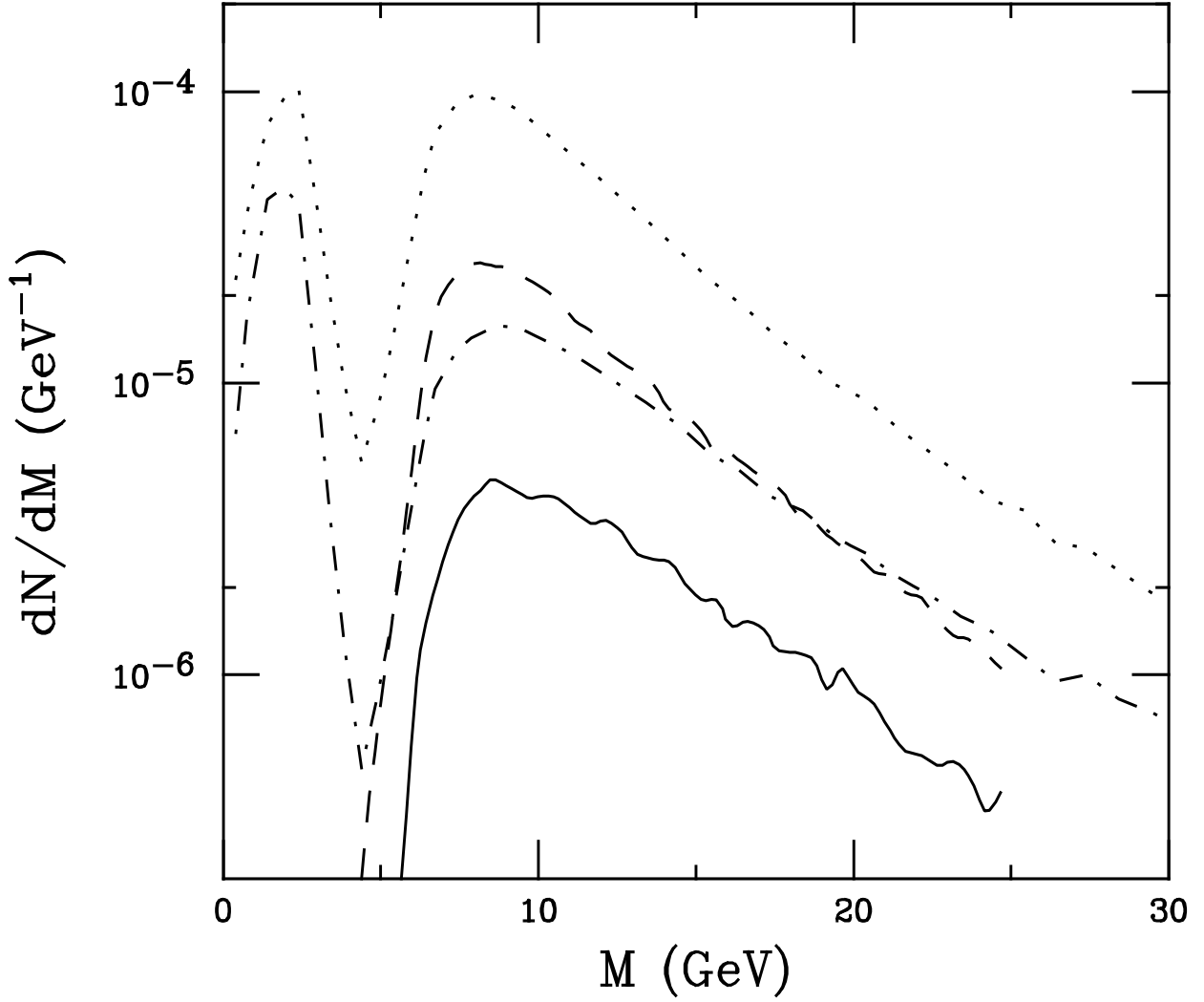}}
\vspace{1cm}
\caption{The dilepton invariant mass distributions in the CMS acceptance.
The dashed and dotted curves are the $D \overline D$ and summed single $B$ and
$B \overline B$ decays respectively without energy loss.  The solid and 
dot-dashed curves are the corresponding results with $dE/dx = -1$ GeV/fm.
}
\label{mcms}
\end{figure}

The CMS \cite{CMS}
muon acceptance is in the range $|\eta| \leq 2.4$ with a lepton
$p_T$ cut of 3 GeV.  After these simple cuts are applied, the results are
shown in Fig.~\ref{mcms} for both $D \overline D$ and $B \overline B$ decays.
Whereas for $M \leq 15$ GeV, the $D \overline D$ decays would dominate those of
$B \overline B$ before the cuts, the measured $B \overline B$ decays are 
everywhere larger
than those from charm mesons both before and after energy 
loss.  The generally larger momentum of muons from $B$ decays and the rather
high momentum cut result in larger acceptance for $B \overline B$ decays.
No $D \overline D$ decay pairs with $M \leq 5$ GeV survive the momentum cut.
A factor of 50 loss in rate at $M \sim 10$ GeV is found before energy loss
when comparing Figs.~\ref{mcms} and \ref{m_cll}.  A loss in rate
by a factor of 100 is obtained when energy loss is included.  The corresponding
acceptance from $B \overline B$ decays is significantly larger, with a loss in
rate of a factor
of $\approx 8$ before energy loss and $\approx 15$ with energy loss.  
Interestingly, the leptons in the decay chain of a single $B$ meson
are energetic enough for both to pass the momentum cut, causing the peak at
$M \sim 2-3$ GeV.  These results suggest that rather than providing an indirect
measurement of the charm cross section, as postulated in \cite{vogt2}, the
dilepton continuum above the $\Upsilon$ family could instead measure the $b
\overline b$ production cross section indirectly.  A comparison with the
spectrum from $pp$ interactions at the same energy would then suggest
the amount of energy loss, $dE/dx$, of the medium.

\begin{figure}[h]
\setlength{\epsfxsize=\textwidth}
\setlength{\epsfysize=0.6\textheight}
\centerline{\epsffile{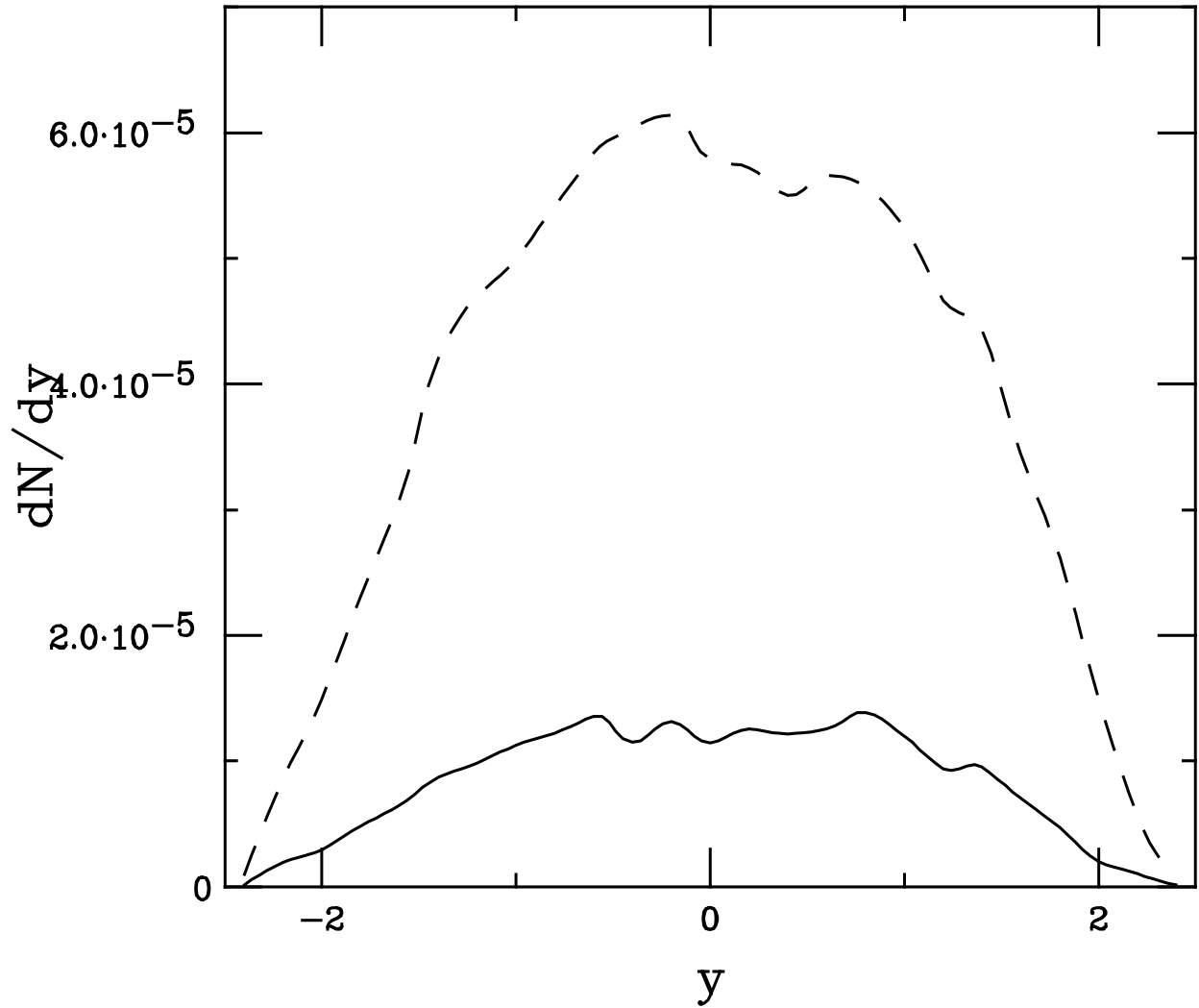}}
\vspace{1cm}
\caption{ The rapidity
distribution of lepton pairs from correlated $D \overline D$ decays,
in the CMS acceptance.  The dashed curve is
without energy loss, the solid curve includes energy loss with $dE/dx =
-1$ GeV/fm.
}
\label{ycms}
\end{figure}

To demonstrate the CMS acceptance, in 
Fig.~\ref{ycms} we show the rapidity distribution of dimuons 
from charm meson decays before and after energy loss.  The broader rapidity
coverage of CMS reduces the effect of energy loss on the dimuon continuum 
relative to ALICE.

\section{Single Leptons from Heavy Quark Decays}
\label{sec-single}

Single leptons from charm decay have been suggested as an indirect measure
of the charm production cross section \cite{tann}. This is possible if
the background leptons from random decays of hadrons such as pions and kaons
can be well understood. 
  
\begin{figure}[h]
\setlength{\epsfxsize=\textwidth}
\setlength{\epsfysize=0.6\textheight}
\centerline{\epsffile{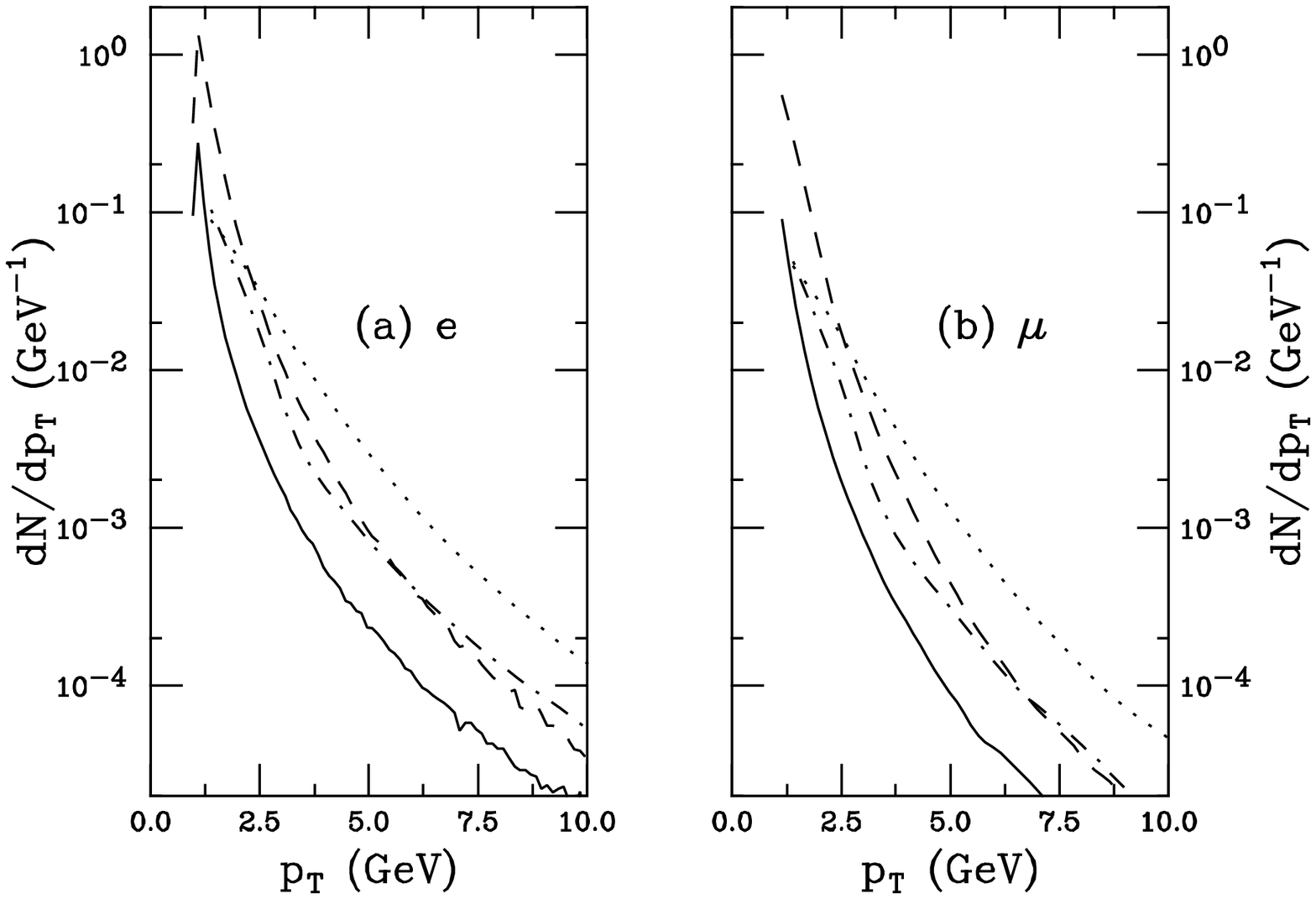}}
\vspace{1cm}
\caption{
The $p_T$ spectrum of single electrons (a) and muons (b) from charm and bottom
decays within the ALICE acceptance.  
The dashed and dotted curves are the $D$ and $B$ meson 
decays respectively without energy loss.  The solid and 
dot-dashed curves are the corresponding results with $dE/dx = -1$ GeV/fm. 
}
\label{ptalice}
\end{figure}

We show the effect of energy loss on single electrons and muons within the
ALICE acceptance in Fig.~\ref{ptalice}.  
Single leptons are not as sensitive to the magnitude of $dE/dx$ as the dilepton
mass spectra.  

Single leptons can be categorized as those from thermalized
heavy quarks and those from heavy quarks energetic enough to
escape after energy loss. 
The former mainly reflects the effective
thermalization temperature while the latter can provide  
us with information on the energy loss.  
Single leptons with energies greater than $1-2$ GeV 
are mainly from energetic heavy quarks and thus are more sensitive to the
energy loss.  Before energy loss, the single leptons from $D$ decays are larger
than those from $b$ hadron decays for $p_T < 2.5$ GeV.  After energy loss, the
$b$ hadron decays dominate the spectra over all $p_T$.

\begin{figure}[h]
\setlength{\epsfxsize=\textwidth}
\setlength{\epsfysize=0.6\textheight}
\centerline{\epsffile{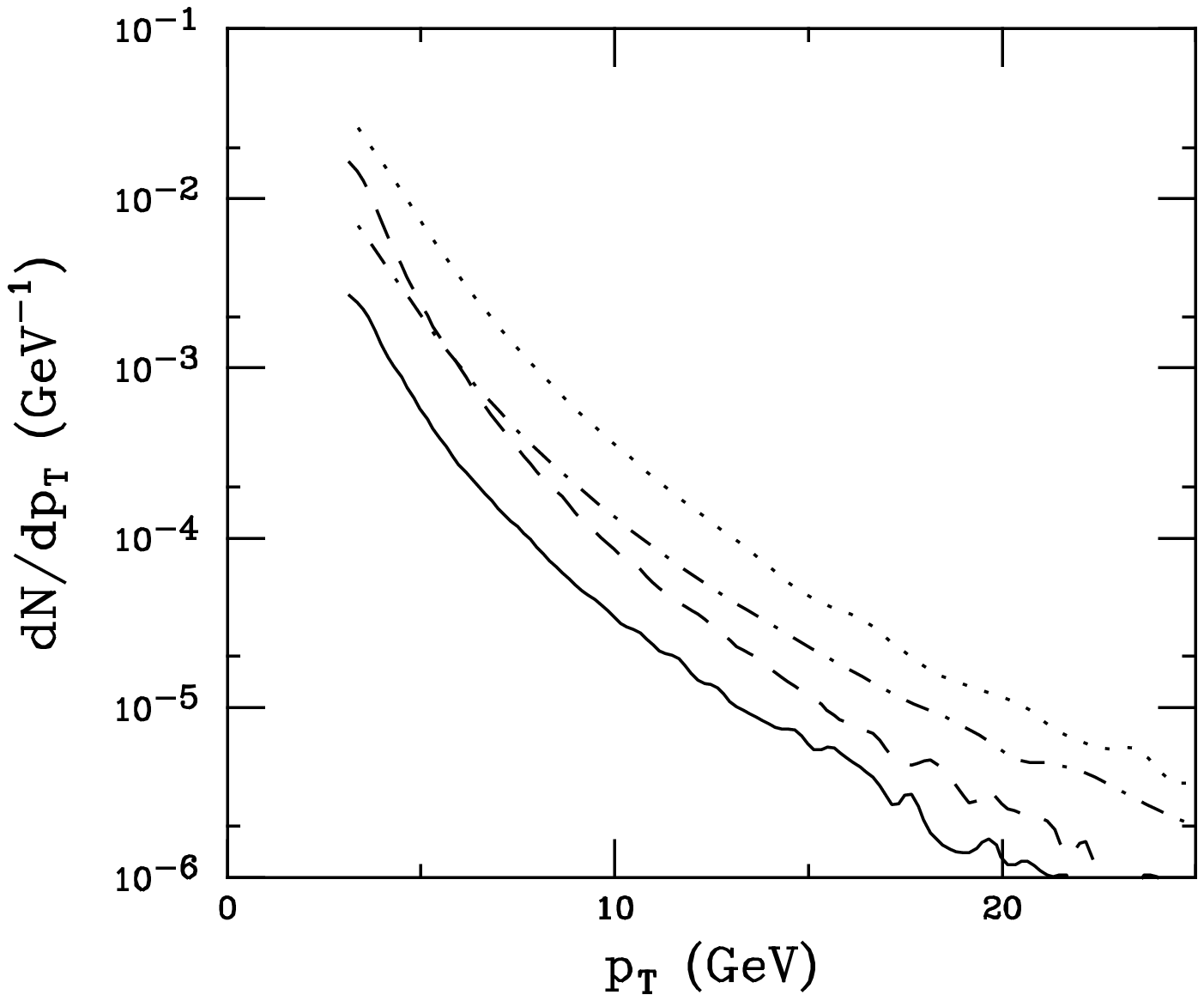}}
\vspace{1cm}
\caption{
The $p_T$ spectrum of single muons from charm and bottom
decays within the CMS acceptance. 
The dashed and dotted curves are the $D$ and $B$ meson 
decays respectively without energy loss.  The solid and 
dot-dashed curves are the corresponding results with $dE/dx = -1$ GeV/fm. 
}
\label{ptcms}
\end{figure}

A comparison of the $p_T$ distributions of single muons in the CMS acceptance
from the decays of $D$ and $B$ mesons can also provide a measure of the $b$
cross section, shown in Fig.~\ref{ptcms}. The muon
$p_T$ distribution is clearly dominated by $B$ decays. 

\section{Summary and Discussion}
\label{sec-summary}

Dileptons from open charm and bottom decays have been calculated for central
Pb+Pb collisions at the LHC including the effect of energy loss on heavy quarks
in dense matter.  

There are a number of uncertainties 
in the model. 
The energy loss is assumed to be constant during the expansion of 
the system and the subsequent drop in the energy density.  This need not
necessarily be the case.
Transverse flow, which could lead to a higher effective temperature, $T$, and
thus enhance the low $p_T$ heavy quark yield and, consequently, the
low invariant mass dilepton yields, is also not included.
However, the qualitative features of the results, such as the clear dominance
of $b \overline b$ decays and the strong suppression due to 
energy loss when $|dE/dx| \geq \langle p_T \rangle/R_A$, 
are not likely to change.  

To determine whether single leptons or dileptons from heavy quark
decays can indeed probe the energy loss, the most important factor is the
magnitude of the random hadron decay background.  
This deserves further study, particularly since high $p_T$ pions will also 
experience quenching effects and be suppressed in high-energy
heavy-ion collisions.  

Acknowledgments: We thank M. Bedjidian, D. Denegri and A. Morsch for
helpful discussions about the ALICE and CMS detectors.

\pagebreak

\pagebreak

\pagebreak

\pagebreak

\pagebreak

\pagebreak

\pagebreak

\pagebreak

\pagebreak

\pagebreak

\pagebreak

\pagebreak

\pagebreak
{}

\end{document}